\documentclass [10pt,compsocconf]{IEEEtran}
\usepackage{algorithm,amsbsy,amsmath,amssymb,epsfig,bbm,mathrsfs,multirow,amsthm,xcolor}
\usepackage[top=0.68in, bottom=1.03in, left=0.64in, right=0.64in]{geometry}
\usepackage{eurosym}
\usepackage{diagbox}
\usepackage{algorithmic}

\usepackage{subcaption,float}
\usepackage[hidelinks]{hyperref}

\usepackage{color}

\usepackage{amsmath}
\DeclareMathOperator*{\argmax}{arg\,max}

\begin{document}
\title{Real-time Cyberattack Detection with Collaborative Learning for Blockchain Networks
\thanks{This work is the output of the ASEAN IVO \url{http://www.nict.go.jp/en/asean_ivo/index.html} project ``Agricultural IoT based on Edge Computing'' and financially supported by NICT \url{http://www.nict.go.jp/en/index.html}. \\Correspondence: Nguyen Linh Trung (linhtrung@vnu.edu.vn).}
}
\author{\IEEEauthorblockN{Tran Viet Khoa$^{1, 2}$, Do Hai Son$^{2, 3}$, Dinh Thai Hoang$^1$, Nguyen Linh Trung$^2$,\\ Tran Thi Thuy Quynh$^2$, Diep N. Nguyen$^1$, Nguyen Viet Ha$^2$, and Eryk Dutkiewicz$^1$ \\}
    	$^1$ School of Electrical and Data Engineering, University of Technology Sydney, Australia \\
	$^2$ AVITECH, University of Engineering and Technology, Vietnam National University, Hanoi, Vietnam \\
    $^3$ Information Technology Institute, Vietnam National University, Hanoi, Vietnam
 \vspace{-5mm}}
\maketitle
\thispagestyle{empty}
\begin{abstract}
With the ever-increasing popularity of blockchain applications, securing blockchain networks plays a critical role in these cyber systems. In this paper, we first study cyberattacks (e.g., flooding of transactions, brute pass) in blockchain networks and then propose an efficient collaborative cyberattack detection model to protect blockchain networks. Specifically, we deploy a blockchain network in our laboratory to build a new dataset including both normal and attack traffic data. The main aim of this dataset is to generate actual attack data from different nodes in the blockchain network that can be used to train and test blockchain attack detection models. We then propose a real-time collaborative learning model that enables nodes in the network to share learning knowledge without disclosing their private data, thereby significantly enhancing system performance for the whole network. The extensive simulation and real-time experimental results show that our proposed detection model can detect attacks in the blockchain network with an accuracy of up to 97\%.            
\end{abstract}

{\it Keywords-} Cyberattack detection, blockchain, distributed machine learning, deep learning, and cybersecurity. 
\section{Introduction}
\label{sec:Introduction}

Blockchain has been emerging as one of the breakthrough technologies for data management in recent years. The most popular blockchain application named Bitcoin~\cite{Bitcoin} was introduced in 2008 with a tight series of blocks in a chain. After data are stored in the chain, they are immutable, secured, and transparent to all mining nodes in a blockchain network. With different outstanding features such as immutability, decentralization, and fault tolerance~\cite{ali2018applications}, a number of blockchain-based applications are introduced to our daily lives such as finance, smart cities, logistics, and healthcare~\cite{ali2018applications}. However, in recent years, these applications have been reported to be a victim of different cyberattacks. For example, in 2020, a cryptocurrency exchange in Singapore named Kucoin was attacked and lost \$281 million~\cite{Top_hack}. Moreover, different reports showed that cyberattacks on various blockchain platforms have been increasing in recent years~\cite{Top_hack}. In different attacks, hackers tried to exploit flaws in blockchain applications (e.g., the processing of a mining node with null transactions or a weak password) to attack the users and affect the working of systems. As a result, it is an urgent need to find effective solutions to detect attacks in blockchain networks.

Machine Learning (ML) has been considered a promising solution that can effectively detect various types of attacks with high accuracy in different networks such as Internet of Things (IoT), edge computing, and cloud computing~\cite{chaabouni2019network}. However, there are only a few works applying ML to deal with cyberattacks in blockchain networks. In particular, in~\cite{kim2021anomaly} the authors first set up an experiment using a real blockchain node. They then use an attack device to perform Eclipse and DoS attacks on that node to collect a cyberattack dataset. The simulation results show that an autoencoder deep learning model can be used to detect attacks with an accuracy of 99\%. In~\cite{liu2021lstm}, the authors propose a model using Long Short-Term Memory Network (LSTM) to better learn the behavior of the network in normal situations using public and private datasets. They also propose to use Conditional Generative Adversarial Networks (CGAN) to create new attack data from normal behavior in the dataset. The simulation results show that the ML methods can detect attacks with an accuracy of 93\%. Moreover, in~\cite{cao2021blockchain}, the authors perform simulations on an Ethereum network and propose a Recurrent Neural Network (RNN) model to detect Link Flood Attack (LFA) which can achieve an accuracy of 99\%. 

Despite the advantage of high accuracy in attack detection, the current ML approaches are still facing a number of challenges. First, lacking synthetic datasets from laboratories is one of the major challenges for all the ML-based solutions~\cite{hassan2021anomaly}. A few recent approaches have been proposed to address this challenge. In~\cite{kim2021anomaly}, the authors collect traffic data from the public Bitcoin network and use them as normal network data. Then, they simulate and generate attack traffic data in the laboratory. Unlike~\cite{kim2021anomaly}, the authors in~\cite{liu2021lstm} use CGAN to create artificial attacks from normal network data. However, using actual blockchain traffic network data as normal data may contain noise data and/or attack data insight. Therefore, such data might make ML models trained improperly, leading to low performance for intrusion detection systems. In addition, the artificial attacks may not truly reflect the properties of actual attacks in practice, leading to ineffective training processes for ML models. Moreover, while blockchain networks are decentralized in nature, all of the current intrusion detection approaches~\cite{kim2021anomaly}\cite{liu2021lstm}\cite{cao2021blockchain} are based on the centralized learning model (i.e., data are connected and trained at a centralized node). This requires distributed nodes to send their local data to the centralized server, causing privacy concerns and excessive network overhead problems.  
\begin{figure*}[t!]
	\centering
	\includegraphics[width=0.75\linewidth]{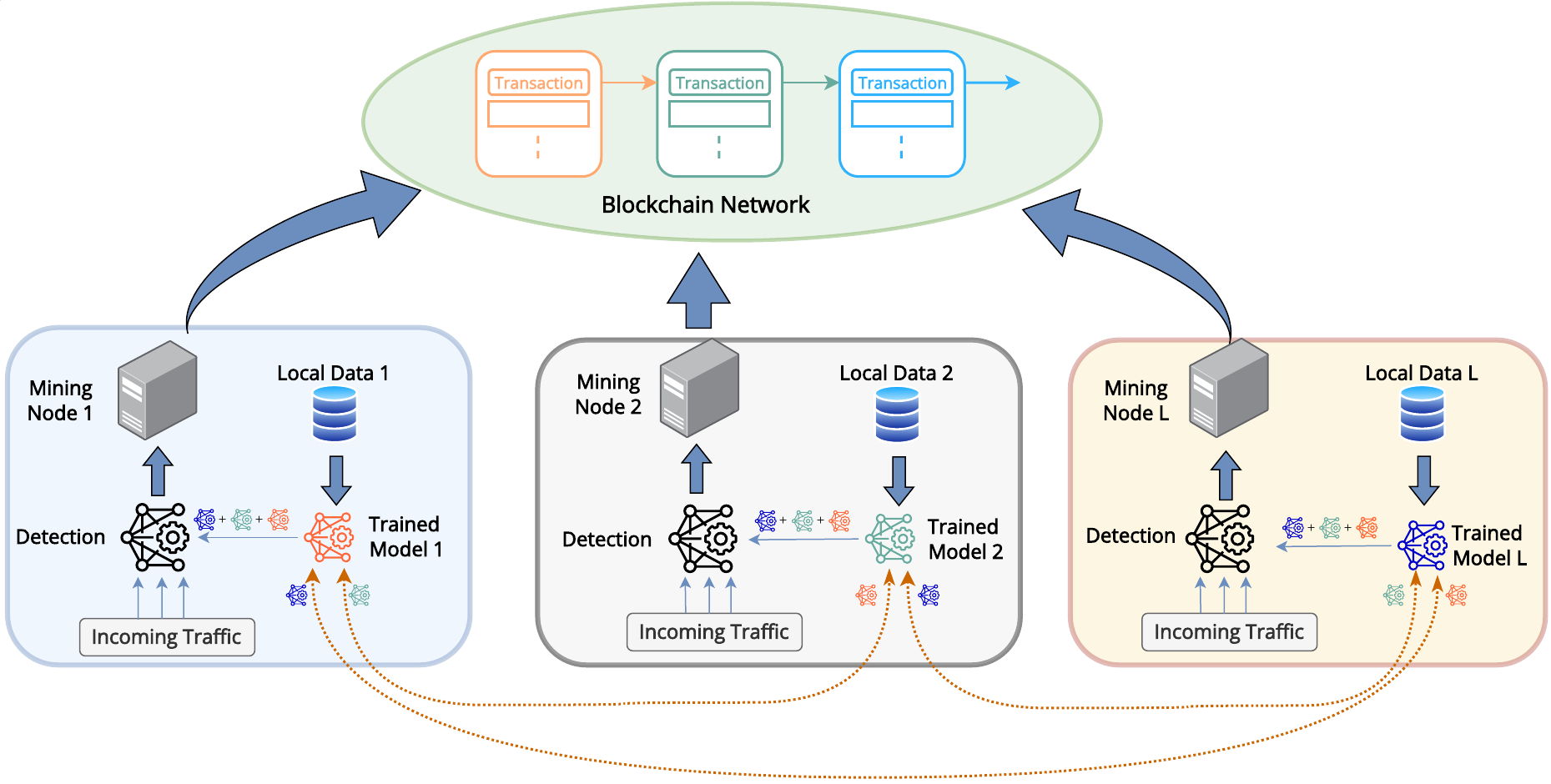}
	\caption{Our proposed collaborative cyberattack detection model. The detection modules are first trained by their local data. They are then used to detect attacks for incoming traffic of blockchain networks before putting them into the mining nodes.}
	\label{fig:sys}
    \vspace*{-0.5cm}
\end{figure*} 

In this paper, to address the problem of lacking data for training ML models in blockchain networks, we first set up a private Ethereum network in our laboratory to build a new dataset, called Blockchain Network Attack Traffic (BNaT), which includes both normal and cyberattack data. We then generate various types of attacks with clean samples (i.e., data without error, duplicate, and corruption), which can be used to train ML models. We then propose a real-time collaborative learning model that can be deployed in a decentralized manner to effectively detect attacks in the blockchain network. Specifically, each node can train its data locally using a deep belief network before sharing the trained model with other nodes in the blockchain network to enhance the accuracy of attack detection. In this case, all the nodes can quickly learn the ``knowledge'' (i.e., via trained models) from other nodes in the network without the need of gathering raw data from all the blockchain nodes in the network for the training process like conventional centralized learning models. Both real-time experiments and extensive simulations show that our proposed approach can outperform other baseline learning models (i.e., centralized learning and individual learning) with an accuracy of up to 97\%.

\section{System Model}
\label{sec:sysmodel}


In a blockchain network, mining nodes (or full nodes) are distributed in various geographical areas. Each mining node takes responsibility for gathering and verifying incoming transactions before the verified transactions can be stored in blocks in the main chain. As a result, mining nodes become the main target of attacks in the blockchain network. It was reported that Kucoin was attacked by Brute Pass and lost \$281 million. In addition, Bitfinex was also attacked by Denial of Service and had to shut down its service to recover. Therefore, detecting attacks to prevent and protect the blockchain network at the mining nodes is a crucial mission for the sustainable development of blockchain networks. 

In this paper, we propose a real-time collaborative cyberattack detection model that can support mining nodes to detect incoming attacks. 
Our proposed collaborative cyberattack detection in the blockchain network is described in Fig.~\ref{fig:sys}. In this model, there are $L$ mining nodes joining the mining processes of a blockchain network. Each mining node uses local private data to train its deep neural learning model (local learning model). The local private data is the labeled data that a mining node possesses for its data training process. However, the amount of data and the number of attacks on each local dataset is usually limited, and thus it may dramatically affect the accuracy of the learning process. Therefore, in this work, we propose a collaborative machine learning model which can help the mining nodes to learn knowledge from other nodes in the networks without the need of sharing their own private datasets. Unlike federated learning models~\cite{Hoang_federatedsurvey} where we need to maintain a centralized node to collect the trained models from all the learning nodes in the network to aggregate and distribute the global model, our proposed collaborative learning model enables the mining nodes to be able to share and learn locally without a need of using the centralized node. In particular, when a mining node obtains its local trained model by training its local data, it will share its model with all other nodes in the network. Once a node receives all the trained models from other nodes in the network, it can aggregate the global model and use this model to update its local trained model. After that, this node will use the updated trained model to continue training its local data. This process is repeated continuously until the iteration reaches a predefined number or the global model is converged. 

\section{Collaborative Cyberattack Detection in Blockchain Networks}
\label{sec:CollaborativeLearning}

\subsection{The Collaborative Classification Learning Model}

Unlike other deep neural networks (e.g., autoencoder deep neural networks), our proposed Deep Belief Network (DBN) uses energy function to optimize its Restricted Boltzmann Machine (RBM) and Gaussian Restricted Boltzmann Machines (GRBM) layers. Thus, the DBN network can be optimized in each layer, and it is more appropriate to classify different types of attacks. 
\begin{figure}[t!]
	\centering
	\includegraphics[width=\linewidth]{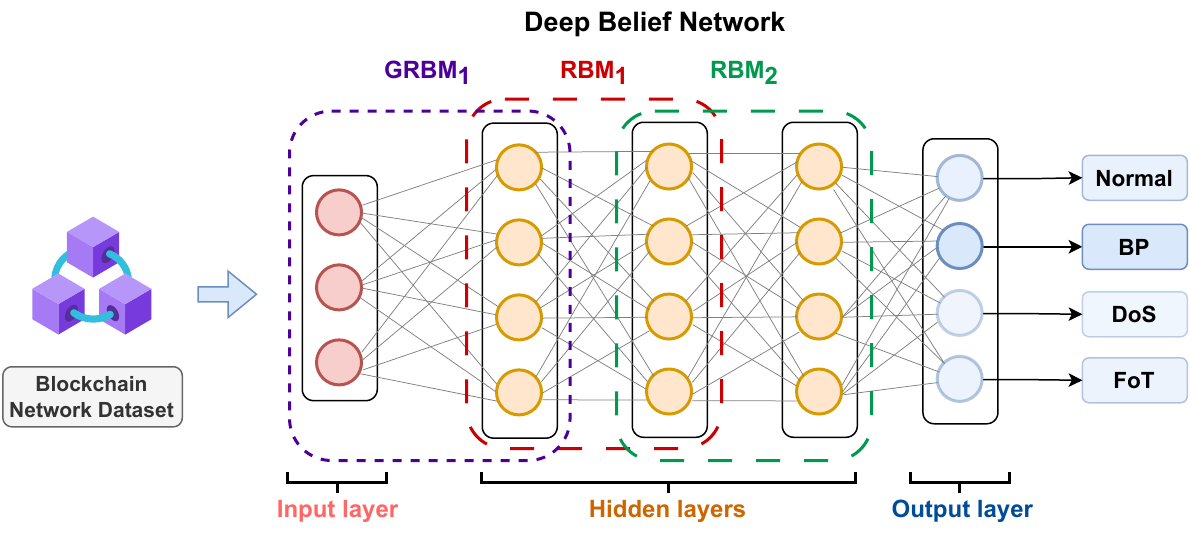}
	\caption{The architecture of a DBN. This architecture includes multiple GRBM and RBM layers for classifying blockchain network traffic.}
	\label{fig:DBN}
    \vspace*{-0.5cm}
\end{figure} 

The architecture of a DBN is illustrated in Fig.~\ref{fig:DBN}. We denote $l \in \{1,\ldots,L\}$ as the number of mining nodes in the blockchain network. In addition, $\boldsymbol{h}^l$ and $\boldsymbol{v}^l$ are the hidden and visible layers of the GRBM and RBM in the neural network of mining node $l$ (MN-$l$), respectively. The numbers of hidden and visible layers of GRBM in this neural network are $G$ and $P$, respectively. We denote $v^l_p$ and $h^l_g$ as the visible layer-$p$ and hidden layer-$g$ of MN-$l$. The energy function of GRBM~\cite{RBM} in the neural network of MN-$l$ is calculated:
\begin{equation}
\begin{aligned}
\label{eqn1}
E^l_{G}(\boldsymbol{v}^l, \boldsymbol{h}^l) = &\sum_{p=1}^P \frac{(v^l_p - b^l_{1,p})^2}{2\gamma_{l,p}^2} - \\&\sum_{p=1}^P\sum_{g=1}^G w^l_{p,g} h^l_g\frac{v^l_p}{\gamma_{l,p}} - \sum_{g=1}^G b^l_{2,g}h^l_g,
\end{aligned}
\end{equation}
where $b^l_{1,p}$ and $b^l_{2,g}$ are the bias parameters, $w^l_{p,g}$ is the weight parameter between the visible layer and hidden layer, and $\gamma_{l,p}$ indicates the standard deviation of the visible layer. From the energy functions in equation~(\ref{eqn1}), we can calculate the probability that $\boldsymbol{v}^l$ is used in the MN-$l$:
\begin{equation}
\begin{aligned}
\label{eqn2}
{\xi}_G^l(\boldsymbol{v}^l) = \frac{\sum_{\boldsymbol{h}^l}e^{-E_G(\boldsymbol{v}^l,\boldsymbol{h}^l)}}{\sum_{\boldsymbol{v}^l,\boldsymbol{h}^l}e^{-E_G(\boldsymbol{v}^l, \boldsymbol{h}^l)}}.
\end{aligned}
\end{equation}

From the equation~(\ref{eqn2}), we can 
calculate the gradient of GRBM layers using the expectation value $\langle.\rangle$~\cite{RBM}:
\begin{equation}
\begin{aligned}
\label{eqn3}
\nabla {\varphi}^l_G
= \sum_{p=1}^{P}\sum_{g=1}^{G} \nabla {\varphi}^l_{G,p,g},
\end{aligned}
\end{equation}
where ${\varphi}^l_{G,p,g}$ is the gradient of a GRBM layer. This gradient can be calculated:
\begin{equation}
\begin{aligned}
\label{eqn4}
\nabla {\varphi}^l_{G,p,g} &= \frac{\partial\log {\xi}_G^l(\boldsymbol{v}^l)}{\partial w^l_{p,g}} \\
&= \langle{ \frac{1}{\gamma_{l,p}} v^l_p h^l_g\rangle}_{dataset} - \langle{ \frac{1}{\gamma_{l,p}} v^l_p h^l_g\rangle}_{model}.
\end{aligned}
\end{equation}

After that, we denote the number of hidden and visible layers of RBM in the neural network of MN-$l$ as $G^*$ and $P^*$, respectively. We can calculate the energy function of RBM~\cite{RBM} in the neural network of MN-$l$: 
\begin{equation}
\begin{aligned}
\label{eqn5}
E^l_{R}(\boldsymbol{v}^l, \boldsymbol{h}^l) = &- \sum_{p=1}^{P^*} b^l_{1,p} v^l_p - \\&\sum_{p=1}^{P^*}\sum_{g=1}^{G^*} w^l_{p,g} v^l_p h^l_g - \sum_{g=1}^{G^*} b^l_{2,g}h^l_g.
\end{aligned}
\end{equation}

Similar to GRBM, we can calculate the gradient of RBM layers in the MN-$l$ as follows:
\begin{equation}
\begin{aligned}
\label{eqn6}
\nabla {\varphi}^l_R
= \sum_{p=1}^{P^*}\sum_{g=1}^{G^*} \nabla {\varphi}^l_{R,p,g},
\end{aligned}
\end{equation}
where
\begin{equation}
\begin{aligned}
\label{eqn7}
\nabla {\varphi}^l_{R,p,g} &= \langle{v^l_p h^l_g\rangle}_{dataset} - \langle{ v^l_p h^l_g\rangle}_{model}.
\end{aligned}
\end{equation}


After processing at the GRBM and RBM layers, the last layer (i.e., the output layer) is used to process data and present the results. To do this, we use the softmax function at the output layer to classify the network behavior based on the processed data after GRBM and RBM. We denote $\boldsymbol{X}_l^*,~\boldsymbol{W}_l^*,~\boldsymbol{b}_l^*$ as the output data of RBM and GRBM layers, the weight matrix and bias between the last hidden layer and output layer at MN-$l$, respectively. Additionally, we denote $u=\{0, ..., U\}$ as the group number of the output. The probability for an output $Y$ to be classified in group $u$ of the MN-$l$:
\begin{equation}
\begin{aligned}
\label{eqn8}
{\xi}^*_l(Y = u|\boldsymbol{X}_l^*,\boldsymbol{W}_l^{*},\boldsymbol{b}_l^{*}) &= softmax(\boldsymbol{W}_l^{*},\boldsymbol{b}_l^{*}), \\
\end{aligned}
\end{equation}
and the output prediction $\boldsymbol{Y}_l$ that has probability ${\xi}^*_l$ is:
\begin{equation}
\begin{aligned}
\label{eqn9}
\boldsymbol{Y}_l= \argmax_u [{\xi}^*_l(Y = u|\boldsymbol{X}_l^*,\boldsymbol{W}_l^{*},\boldsymbol{b}_l^{*})].
\end{aligned}
\end{equation}

The gradient of the last hidden layers and the output can be calculated:
\begin{equation}
\begin{aligned}
\label{eqn10}
\nabla {\varphi}^*_l &= \frac{\partial {\xi}^*_l(Y = u|\boldsymbol{X}_l^*,\boldsymbol{W}_l^{*},\boldsymbol{b}_l^{*})}{\partial \boldsymbol{W}_l^{*}}.
\end{aligned}
\end{equation}

The total gradient $\nabla {\varphi}_l$ of the DBN in MN-$l$ can be calculated using the results of equations~(\ref{eqn3}),~(\ref{eqn6}) and~(\ref{eqn10}). 
The total gradient of the DBN in MN-$l$ can be calculated:
\begin{equation}
\begin{aligned}
\label{eqn11}
\nabla {\varphi}_l = \nabla {\varphi}^l_G + \nabla {\varphi}^l_R + \nabla {\varphi}^*_l.
\end{aligned}
\end{equation}

At each iteration, the DBN of MN-$l$ can calculate its gradient $\nabla {\varphi}_l$ and send it to other nodes to calculate the average gradient. When a node receives $(L-1)$ gradients from other nodes, the average gradient can be calculated~\cite{Federated_learning2}:
\begin{equation}
\begin{aligned}
\label{eqn12}
\nabla {\varphi}' = \frac{1}{L} \sum_{l=1}^L \nabla {\varphi}_l.
\end{aligned}
\end{equation}

We denote $\Theta_i$ as the global model at iteration $i$. Here, $\epsilon$ is the learning rate. Based on the results in equation~(\ref{eqn12}), we can calculate the global model at the next iteration:
\begin{equation}
\begin{aligned}
\label{eqn13}
\boldsymbol{\Theta}_{i+1} = \boldsymbol{\Theta}_i + \epsilon \nabla {\varphi}'.
\end{aligned}
\end{equation}

Finally, each mining node can use $\boldsymbol{\Theta}_{i+1}$ as a new global model. The DBN of each mining node then uses this global model to update its parameters. We denote $\boldsymbol{W}^{*}_{Global}$ as the weight parameters between the last hidden layer and the output layer of the global model. The final output of DBN can be calculated:
\begin{equation}
\begin{aligned}
\label{eqn14}
\boldsymbol{Y}_l= \argmax_u [{\xi}^*_l(Y = u|\boldsymbol{X}_l^*,\boldsymbol{W}^{*}_{Global},\boldsymbol{b}_l^{*})],
\end{aligned}
\end{equation}

Based on equation~(\ref{eqn14}), the deep learning model in each mining node can classify the behavior of the network, e.g., normal or a type of attack. In summary, Algorithm~\ref{al:Classification_FL} describes the learning process of our proposed real-time collaborative cyberattack detection model.
\begin{algorithm}[t]
	\algsetup{linenosize=\tiny}
	\caption{Learning process of the proposed model}
	\label{al:Classification_FL}
	\begin{algorithmic}[1]
	    \WHILE{$i$ $\leq$ predefined iterations}
	        \FOR{$\forall l \in L$}
        		\STATE $\boldsymbol{X}_l$ is put into the DBN of MN-$l$ to create $\boldsymbol{Y}_l$. 
        		\STATE Calculate $\nabla {\varphi}_l$ and send it to other mining nodes.
		    \ENDFOR
		    \STATE Each node receives $L-1$ trained models from other mining nodes.
            \STATE Each node calculates $\nabla {\varphi}'$. 
		    \STATE {$i=i+1$.}
		    \STATE Each node calculates a new global model $\boldsymbol{\Theta}_{i+1}$ and updates its local model.
		\ENDWHILE
		\STATE Each node uses the global model $\boldsymbol{\Theta}_{i+1}$ to predict $\boldsymbol{Y}_l$ from incoming transactions $\boldsymbol{X}_l$.
	\end{algorithmic}
\end{algorithm}
 
\subsection{Evaluation Methods}

In this paper, we use the confusion matrix~\cite{confusion_matrix1} to evaluate our proposed model. The accuracy, precision, and recall of the confusion matrix have been widely used to evaluate the performance of models, especially deep learning models because it provides a comprehensive view to evaluate the output results with the labels. We denote TP as “True Positive”, FP as “False Positive”, TN as “True Negative”, and FN as “False Negative”. Given $U$ number of classes of network behavior (normal and different types of attacks), the accuracy of systems can be calculated as follows:
\begin{equation}
\begin{aligned}
\label{eqn15}
	ACC=\frac{1}{U}\sum_{u=1}^{U} \frac{\mbox{TP}_u+\mbox{TN}_u}{\mbox{TP}_u+\mbox{TN}_u+\mbox{FP}_u+\mbox{FN}_u}.
\end{aligned}
\end{equation}

\section{Experiment Setup and Attack Dataset Collection}

\subsection{Experiment Setup}

\begin{figure}
    \centering
    \includegraphics[width=.9\linewidth]{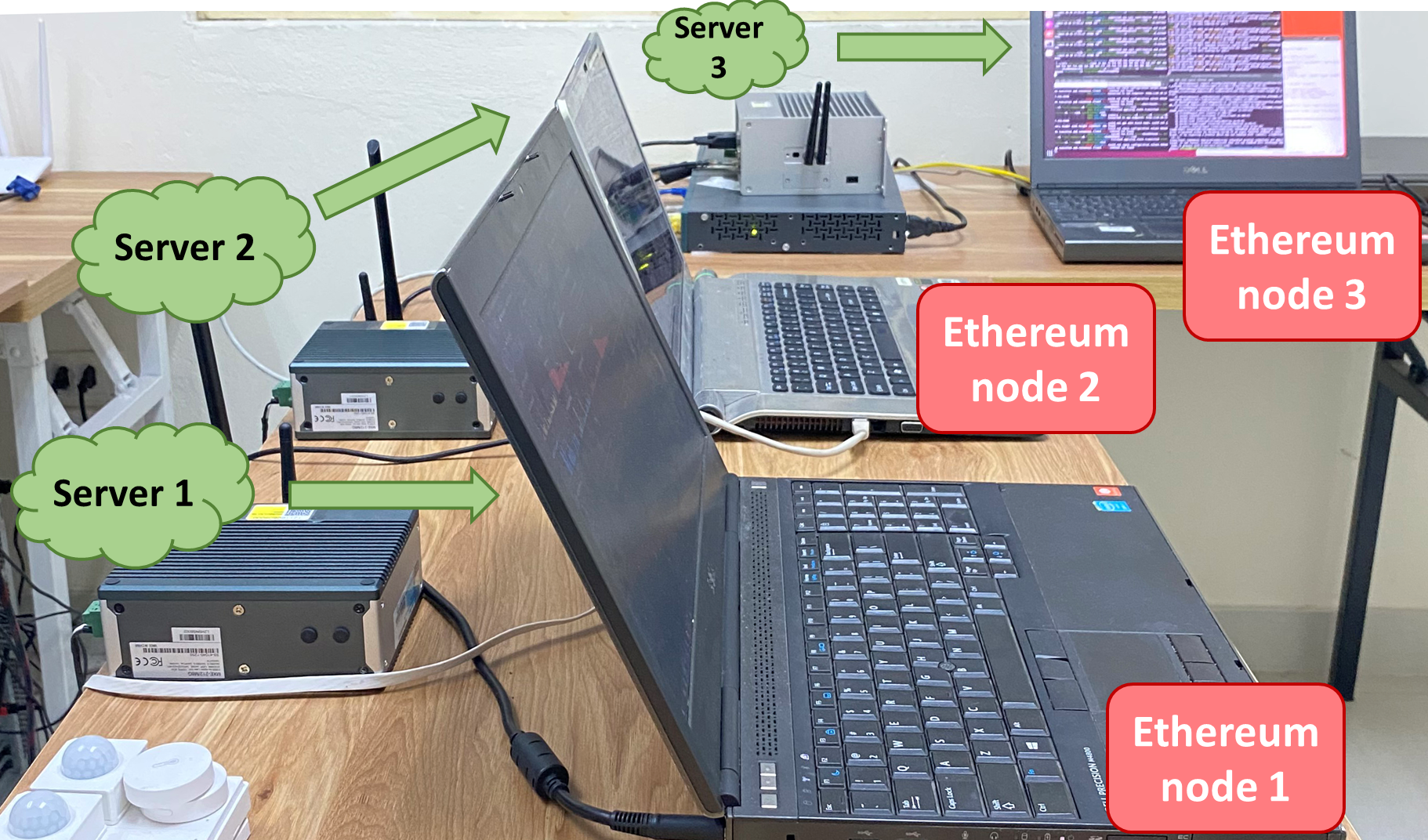}
    \caption{Experiment setup in our laboratory. This experiment includes three Ethereum nodes and three servers in a network.}
    \label{fig:real}
    \vspace*{-0.5cm}
\end{figure}

Real-world cyberattacks at the network layer of blockchain networks usually try to disrupt peer-to-peer connections or target the weakness of consensus mechanisms to steal users' digital assets (i.e., coins, tokens, NFT). Hence, our experiments focus on three kinds of cyberattacks that have been reported for severe financial loss in the laboratory environment as follows:
\begin{itemize}
    \item \textit{Brute-force Password (BP)}: Blockchain wallets are usually secured with users' passwords. Hackers can use the traditional cyberattack as BP to scan the wallet password. In detail, if users' wallet passwords are not strong enough (e.g., not including upper and special characters), hackers can quickly retry the password until they find the correct one. KuCoin~\cite{Top_hack} lost up to \$281 million due to such attacks in 2020. 

    \item \textit{Denial of Service (DoS)}: Blockchain nodes can be easily attacked by DoS attacks. When hackers send a massive amount of traffic to target blockchain nodes from botnets, these nodes will not be able to receive blockchain transactions, mining or even be stopped. In 2018, two DoS attacks happened consecutively, and they caused Bitfinex~\cite{Bitfinex} to shut down temporarily.

    \item \textit{Flooding of Transactions (FoT)}: Blockchain blocks can only contain a certain number of transactions~\cite{floodtran}. Hackers can send a large number of null transactions to slow down the confirmation of honest transactions. In the real world, 115,000 transactions were unconfirmed in the Bitcoin network in 2017, which led to \$700 million being lost~\cite{saad2020exploring}.
\end{itemize}

In order to capture the dataset, we set up a private Ethereum blockchain network in our laboratory as shown in Fig.~\ref{fig:real}. This network includes three $Geth$ clients~\cite{Go-Ethereum} as Ethereum blockchain nodes and three servers. The servers receive transactions and send them to the Ethereum blockchain nodes for the mining process. We also use a server to perform attacks on this Ethereum blockchain network to collect attack data. By using collected data for training, our model can be extended to use with other types of attacks.  

\subsection{Attack Dataset Collection}

\begin{figure}[t]
    \centering
    \begin{subfigure}{0.4\linewidth}
        \includegraphics[width=\linewidth]{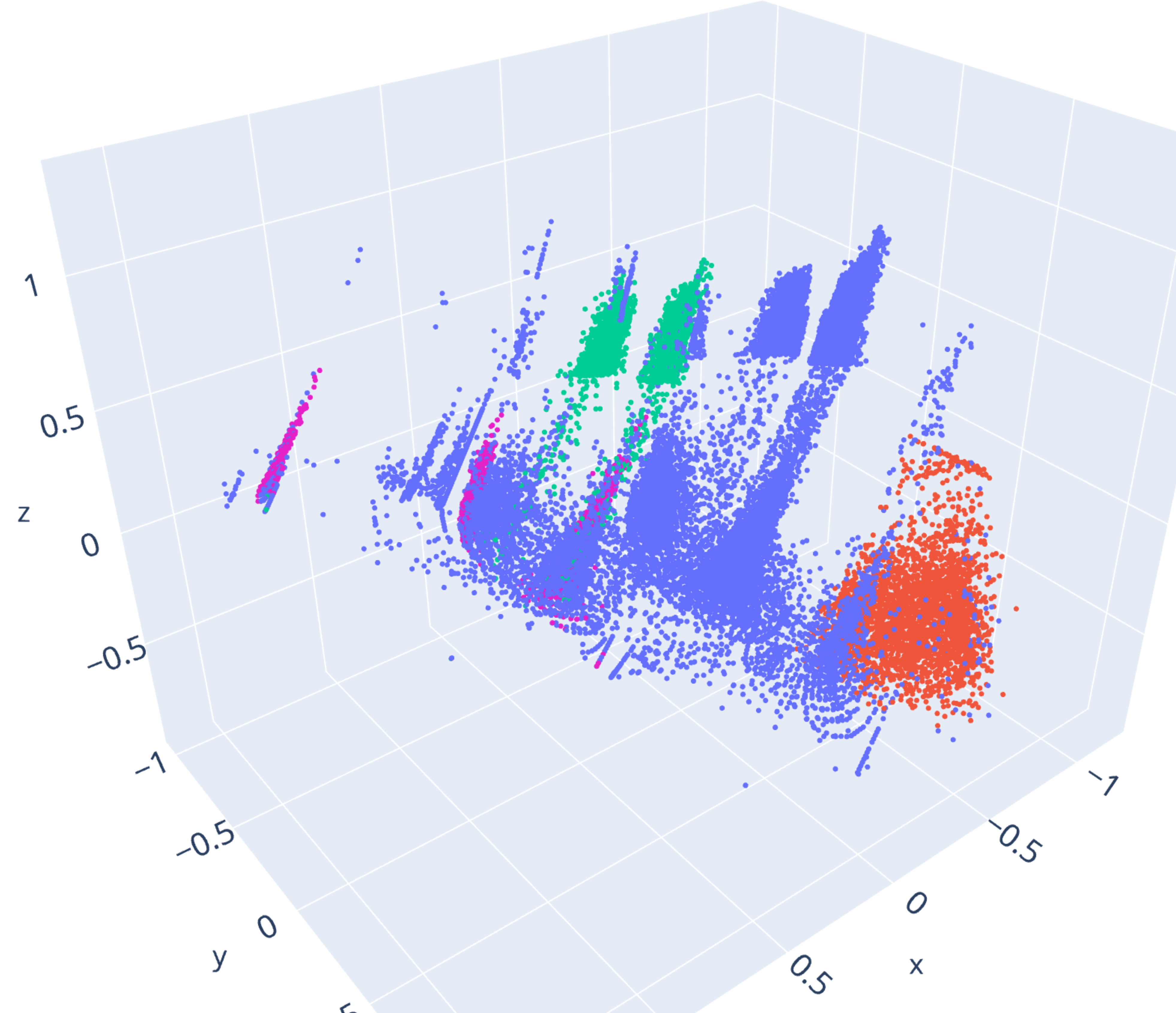}
        \caption{Dataset at node 1}
        \label{fig:dataset_visu_1}
    \end{subfigure}
    \hfill
    \begin{subfigure}{0.4\linewidth}
        \includegraphics[width=\linewidth]{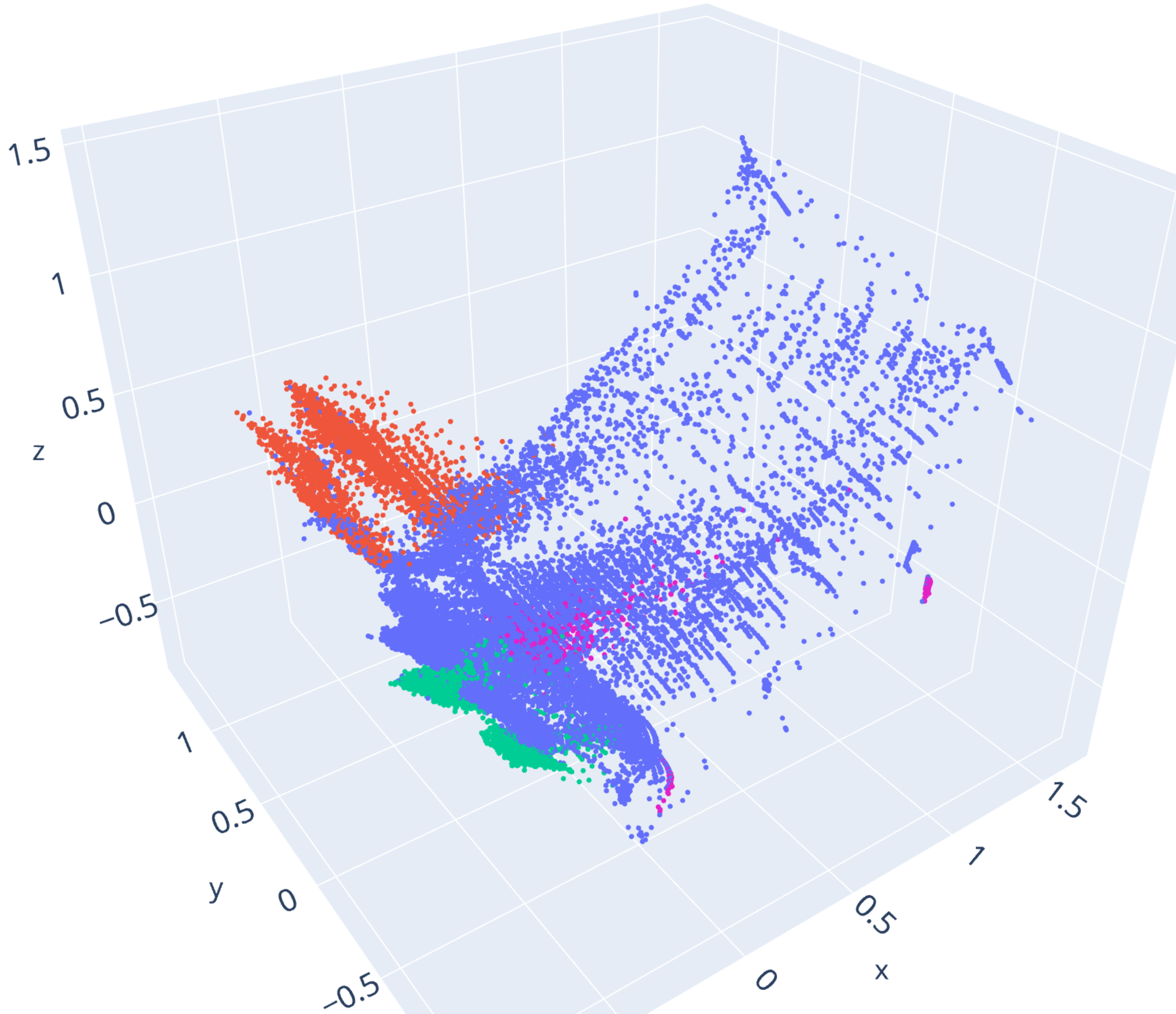}
        \caption{Dataset at node 2}
        \label{fig:dataset_visu_2}
    \end{subfigure}
    \hfill
    \begin{subfigure}{0.4\linewidth}
        \includegraphics[width=\linewidth]{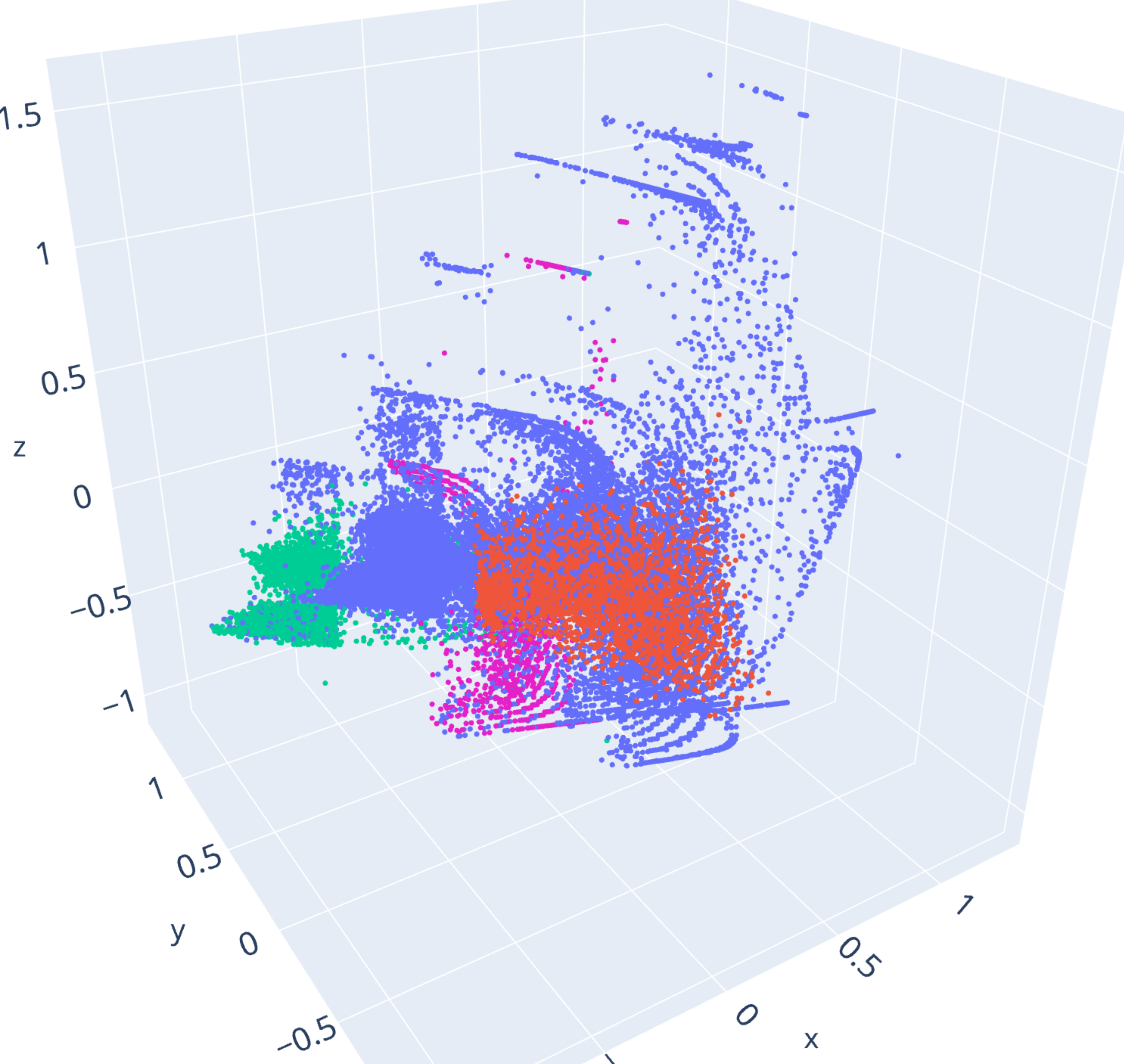}
        \caption{Dataset at node 3}
        \label{fig:dataset_visu_3}
    \end{subfigure}
    \hfill
    \begin{subfigure}{0.4\linewidth}
        \includegraphics[width=\linewidth]{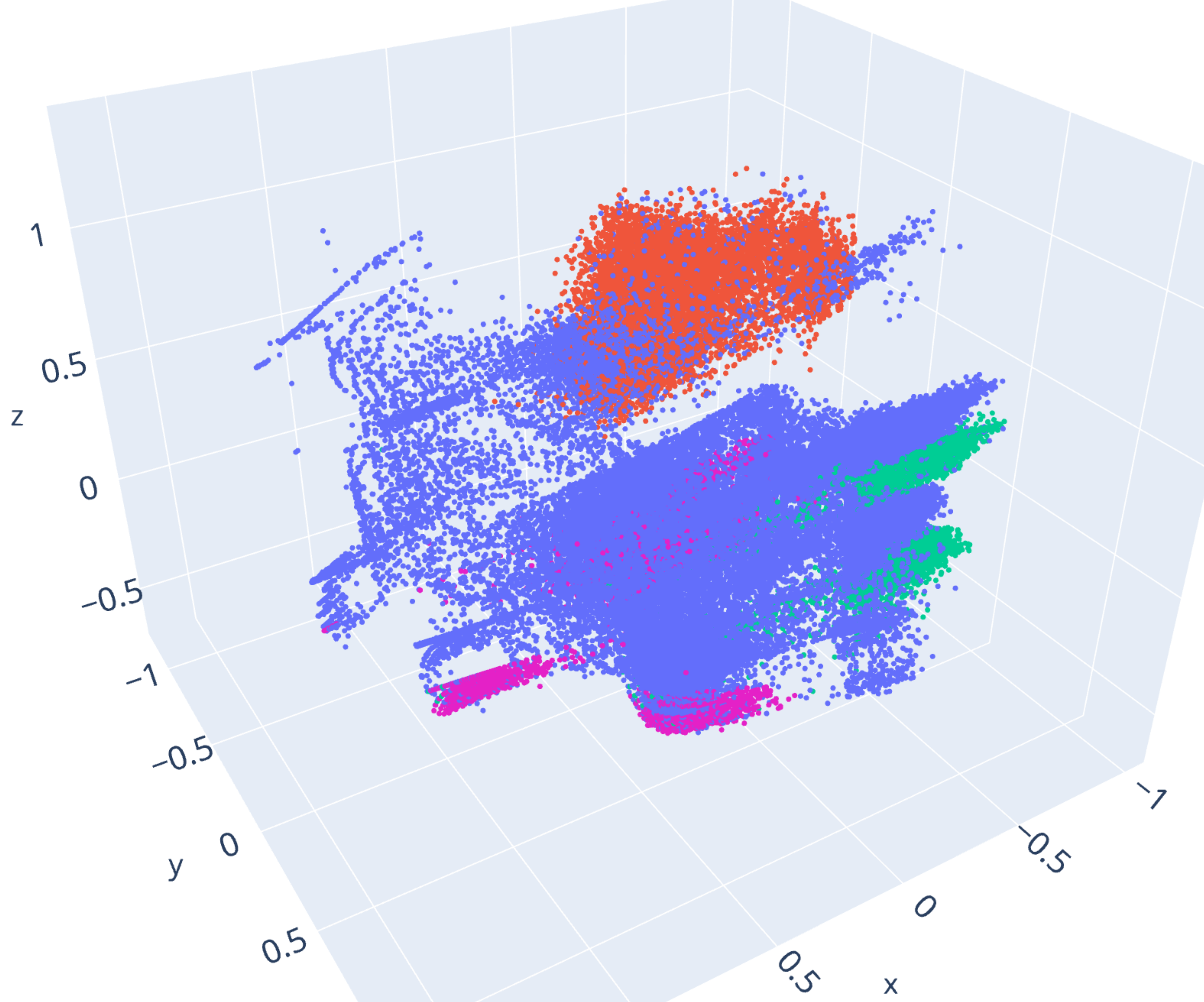}
        \caption{Combine data from 3 nodes}
        \label{fig:dataset_visu_4}
    \end{subfigure}
    \hfill
    \begin{subfigure}{\linewidth}
        \centering
        \includegraphics[width=0.7\linewidth]{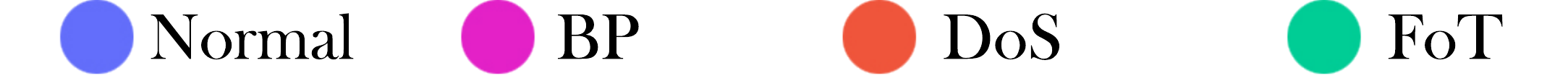}
    \end{subfigure}
     \vspace*{-0.3cm}
    \hfill
    \caption{Visualization using PCA for collected datasets.}
    \label{fig:dataset_visu}
    \vspace*{-0.5cm}
\end{figure}

At each node, network traffic data of four states are captured, which include the normal state (Class 1), and three attack states, i.e., BP attack (Class 2), DoS attack (Class 3), and FoT attack (Class~4). In a normal network, traffic data often scatters on different \textit{port numbers} of the node's operating system. However, the traffic of Ethereum is always on fixed $ports$ that were set up when initializing the Ethereum node. 
Table~\ref{tab:dataset} shows the number of samples in each class of our dataset. Fig.~\ref{fig:dataset_visu} shows the visualization of the collected dataset using Principal Component Analysis (PCA) with the three most important features. We can observe in this figure that the visualization of DoS attack samples is quite separated from the other states of the network (i.e., normal, BP, FoT). However, in the 3D view as shown in Fig.~\hyperref[fig:dataset_visu_4]{4(d)}, when we combine datasets from all the nodes, all attack states are overlapped with normal state points. This makes more challenges in classifying normal and attack traffic data when both types of data are mixed.
\begin{table}[!t]
    \centering
    \caption{The number of samples of dataset (BNAT) collected in our laboratory.}
    \label{tab:dataset}
    \begin{tabular}{|l|c|c|c|c|} 
        \hline
        \multicolumn{1}{|c|}{\textbf{\diagbox[width=\dimexpr \textwidth/6-\tabcolsep\relax, height=1cm]{Ethereum node}{Class}}} & \begin{tabular}[c]{@{}c@{}}\textbf{Normal}\\(Class 1)~\end{tabular} & \begin{tabular}[c]{@{}c@{}}\textbf{BP}\\(Class 2)~\end{tabular} & \begin{tabular}[c]{@{}c@{}}\textbf{DoS}\\(Class 3)~\end{tabular} & \begin{tabular}[c]{@{}c@{}}\textbf{FoT}\\(Class 4)\end{tabular} \\ 
        \hline
        Node 1 (samples) & 30,000 & 3,000 & 3,000 & 3,000 \\ 
        \hline
        Node 2 (samples) & 30,000 & 3,000 & 3,000 & 3,000 \\ 
        \hline
        Node 3 (samples) & 30,000 & 3,000 & 3,000 & 3,000 \\
        \hline
        Total (samples) & 90,000 & 9,000 & 9,000 & 9,000 \\
        \hline
    \end{tabular}
\end{table}

\subsection{Simulation Results}

After collecting the dataset, we implement experiments to evaluate the performance of the proposed collaborative cyberattack detection model in comparison with other learning models. In these experiments, each network has its own collected dataset, and this dataset is separated into a training group and a testing group. Besides, each network also has a DBN to classify the dataset. However, this DBN can work in different scenarios of the experiments. In detail, we consider the following schemes for evaluation:

\begin{itemize}
    \item \textit{Centralized Learning Model (CLM)}: This is an upper-bound baseline solution when we assume that there exists one centralized node that can gather training data of all nodes in the blockchain network. After that, to make fair comparisons, we use a DBN to train all the data at the centralized node. The trained DBN is then used to evaluate the accuracy in detecting attacks. This will be used as an upper-bound benchmark to compare with our proposed approach. 
    
    \item \textit{Local Learning Model (LLM)}: In this baseline, we assume that the nodes are not cooperative, and they train their DBN locally based on their own data. The trained DBNs are then used to evaluate the accuracy in detecting attacks.  
\end{itemize}

\subsubsection{Convergence analysis}

\begin{figure}[t!]
	\centering
	\includegraphics[width=1\linewidth]{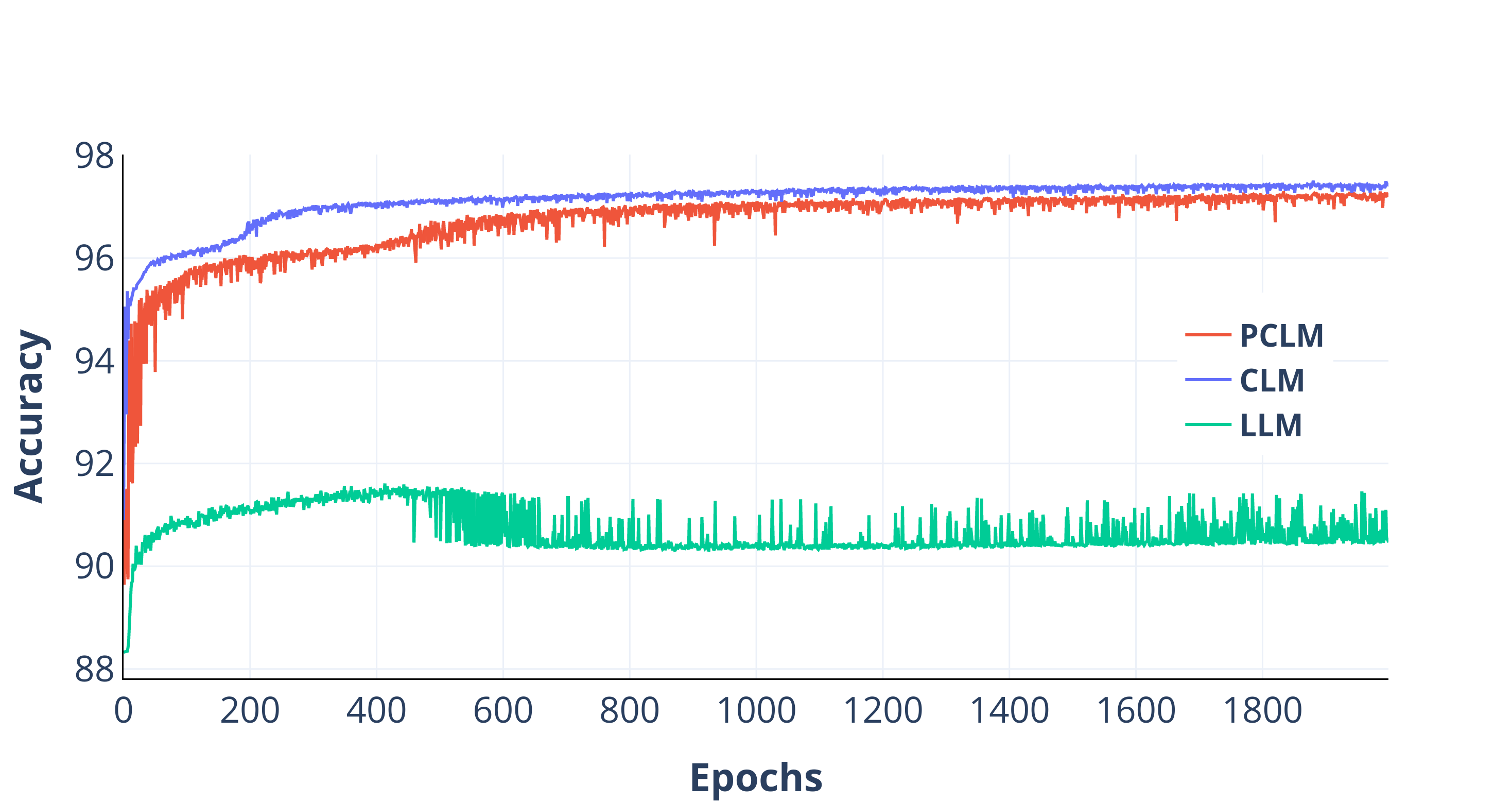}
	\caption{The convergence of learning models in three schemes.}
	\label{fig:coverage}
    \vspace*{-0.65cm}
\end{figure}

Fig.~\ref{fig:coverage} describes the convergence of learning rates of three different schemes, i.e., our Proposed Collaborative Learning Model (PCLM), Centralized Learning Model (CLM), and Local Learning Model (LLM). We can observe that the CLM can quickly reach convergence after only 400 epochs. In addition, the accuracy of CLM is much higher than that of the ILM, i.e., 96.91\% v.s. 92.9\%, respectively. The reason is that the CLM has more training data than that of the LLM. Our proposed model converges after 700 epochs, which is a bit longer than those of the other models because it needs time to exchange knowledge among the nodes. Interestingly, although it does not need to maintain a centralized node to collect all the data in the network to train, its accuracy in detecting attacks is very close to that of the CLM, i.e., 96.72\% vs 96.91\%, respectively. 

\subsubsection{Performance evaluation}
\begin{table*}[t]
\centering
\caption{Simulation results.}
    \label{tab:simu_result}
    \resizebox{\linewidth}{!} {
    \begin{tabular}{|l|c|c|c|c|c|c|c|c|c|c|c|c|c|c|c|} 
        \hline
        \multirow{3}{*}{\textbf{\textbf{\diagbox[width=\dimexpr \textwidth/8-\tabcolsep\relax, height=0.9cm]{}{Model}}}} & \multicolumn{6}{c|}{\textbf{2 Nodes}} & \multicolumn{9}{c|}{\textbf{3 Nodes}} \\ 
        \cline{2-16}
         & \multicolumn{2}{c|}{\textbf{PCLM}} & \multicolumn{2}{c|}{\textbf{CLM}} & \multicolumn{2}{c|}{\textbf{LLM}} & \multicolumn{3}{c|}{\textbf{PCLM}} & \multicolumn{3}{c|}{\textbf{\textbf{CLM}}} & \multicolumn{3}{c|}{\textbf{\textbf{LLM}}} \\ 
        \cline{2-16}
         & \multicolumn{1}{l|}{\textbf{\textbf{Node 1}}} & \multicolumn{1}{l|}{\textbf{\textbf{Node 2}}} & \multicolumn{1}{l|}{\textbf{\textbf{Node 1}}} & \multicolumn{1}{l|}{\textbf{\textbf{Node 2}}} & \textbf{Node 1} & \textbf{Node 2} & \multicolumn{1}{l|}{\textbf{\textbf{Node 1}}} & \multicolumn{1}{l|}{\textbf{\textbf{\textbf{\textbf{Node 2}}}}} & \multicolumn{1}{l|}{\textbf{\textbf{\textbf{\textbf{Node 3}}}}} & \multicolumn{1}{l|}{\textbf{\textbf{Node 1}}} & \multicolumn{1}{l|}{\textbf{\textbf{\textbf{\textbf{Node 2}}}}} & \multicolumn{1}{l|}{\textbf{\textbf{\textbf{\textbf{Node 3}}}}} & \textbf{Node 1} & \textbf{\textbf{Node 2}} & \textbf{\textbf{Node 3}} \\ 
        \hline
        \textbf{Accuracy} & \textbf{96.679} & \textbf{96.722} & 96.756 & 96.919 & 90.927 & 92.932 & \textbf{97.081} & \textbf{97.056} & \textbf{97.248} & 97.432 & 97.350 & 97.466 & 90.462 & 91.939 & 90.244 \\ 
        \hline
        \textbf{Precision} & \textbf{93.271} & \textbf{93.287} & 93.117 & 93.471 & 68.904 & 82.245 & \textbf{94.037} & \textbf{93.973} & \textbf{94.318} & 94.805 & 94.584 & 94.845 & 68.296 & 80.185 & 74.737 \\ 
        \hline
        \textbf{Recall} & \textbf{93.359} & \textbf{93.444} & 93.513 & 93.838 & 81.855 & 85.863 & \textbf{94.162} & \textbf{94.111} & \textbf{94.496} & 94.863 & 94.701 & 94.932 & 80.923 & 83.860 & 80.487 \\
        \hline
    \end{tabular}
    }
\end{table*}

\begin{table*}[!h]
\centering
\caption{Real-time experimental results.}
\label{tab:real_result}
\resizebox{.7\linewidth}{!} {
\begin{tabular}{|l|c|c|c|c|c|c|c|c|c|c|} 
\hline
\multirow{3}{*}{\textbf{\diagbox[width=\dimexpr \textwidth/8-\tabcolsep\relax, height=0.9cm]{}{Model}}}  & \multicolumn{4}{c|}{\textbf{2 Nodes}} & \multicolumn{6}{c|}{\textbf{3 Nodes}} \\ 
\cline{2-11}
 & \multicolumn{2}{c|}{\textbf{PCLM}} & \multicolumn{2}{c|}{\textbf{CLM}} & \multicolumn{3}{c|}{\textbf{PCLM}} & \multicolumn{3}{c|}{\textbf{CLM}} \\ 
\cline{2-11}
  & \textbf{Node 1} & \textbf{\textbf{Node 2}} & \textbf{\textbf{Node 1}} & \textbf{\textbf{\textbf{\textbf{Node 2}}}} & \textbf{\textbf{Node 1}} & \textbf{\textbf{\textbf{\textbf{Node 2}}}} & \textbf{\textbf{\textbf{\textbf{Node 3}}}} & \textbf{\textbf{Node 1}} & \textbf{\textbf{\textbf{\textbf{Node 2}}}} & \textbf{Node 3} \\ 
\hline
\textbf{Accuracy} & \textbf{93.354} & \textbf{95.507} & 93.317 & 92.795 & \textbf{94.565} & \textbf{96.012} & \textbf{93.701} & 94.279 & 96.164 & 92.887 \\ 
\hline
\textbf{Precision} & \textbf{85.657} & \textbf{91.852} & 90.840 & 89.711 & \textbf{88.127} & \textbf{93.515} & \textbf{79.683} & 87.916 & 92.787 & 76.194 \\ 
\hline
\textbf{Recall} & \textbf{86.709} & \textbf{91.015} & 86.634 & 85.589 & \textbf{89.130} & \textbf{92.023} & \textbf{87.401} & 88.558 & 92.328 & 85.774 \\
\hline
\end{tabular}
}
\vspace*{-0.5cm}
\end{table*}

Table~\ref{tab:simu_result} describes the simulation results with two and three nodes in the blockchain network. 
In this table, we evaluate the performance based on the confusion matrix, i.e., accuracy, precision, and recall. Overall, the accuracy, precision, and recall of both the PCLM and CLM are very close and much higher than those of the LLM. When the number of mining nodes increases from two to three, the accuracy of LLM seems unchanged and keeps stable at around 91\%. However, there still has a big gap in the precision of three nodes from 68-80\%. In contrast, the accuracy, precision, and recall of PCLM and CLM are still stable and keep high at about 97.24\%, 94.31\%, and 94.49\%, respectively. These results show that PCLM can leverage the learning knowledge of the neural network in other networks to improve the accuracy of cyberattack detection without sharing private data in the network.


\subsection{Experimental Results}

In this section, we implement experiments to evaluate the real-time cyberattack detection of PCLM and CLM. To do this, we use the learning model after training with PCLM and CLM and run them on the mining nodes to detect attacks in real-time. Our extraction tool continuously extracts the blockchain network traffic into data features and labels them every 2 seconds. The experiments are set up in two cases: two and three mining nodes. Table~\ref{tab:real_result} describes the real-time experimental results. 
In general, we can observe that both PCLM and CLM can detect cyberattacks in the blockchain network in real-time with high accuracy at up to 95.5\%. However, we can observe that in the case of 2 nodes, the accuracy of PCLM is higher than that of CLM about 2.5\% at Node 2. Similarly, in the case of 3 nodes, the precision of PCLM is higher than that of CLM about 3.2\% at Node 3. These results demonstrate that even though our PCLM does not need to gather data from the other nodes in the blockchain network, it still has better performance than that of the CLM which needs to collect and train data from all the nodes in the whole network.             

\section{Conclusion}
In this paper, we built a cyberattack dataset of a blockchain network and proposed a collaborative cyberattack detection model for a blockchain network. We first implemented the experiments in our laboratory to build the blockchain network and collect data (both normal and attack traffic data). After that, we proposed a real-time collaborative learning model that can efficiently detect attacks in a blockchain network. Our proposed model could not only protect the privacy of data in participating networks but also enhance the accuracy of cyberattack detection in a blockchain network. Both simulations and real-time experimental results showed the outperformance of our proposed model in comparison with other methods. In the future, we can expand the scope of our work by studying more types of attacks, and find more effective learning models to enhance the accuracy in detecting attacks in blockchain networks.     


\bibliography{reference2, related_works}
\bibliographystyle{IEEEtran}

	
\end{document}